\documentclass{jpp}
\usepackage{graphicx}
\usepackage{epstopdf, epsfig}
\usepackage{hyperref}
\usepackage{amssymb}
\usepackage{amsmath}
\usepackage{xcolor}
\usepackage{cancel}

\title{On the importance of slow ions in the kinetic Bohm criterion} 
\shortauthor{A. Geraldini, S. Brunner}

\shorttitle{On the importance of slow ions in the kinetic Bohm criterion}
\author{Alessandro Geraldini\corresp{\email{alessandro.geraldini@epfl.ch}}, Stephan Brunner
  }

\affiliation{Swiss Plasma Center (SPC), \'Ecole Polytechnique F\'ed\'erale de Lausanne (EPFL), CH-1015 Lausanne, Switzerland}

\begin{document}
  \maketitle

\begin{abstract}
Between a plasma and a solid target lies a positively charged sheath of several Debye lengths $\lambda_{\rm D}$ in width, typically much smaller than the characteristic length scale $L$ of the main plasma.
This scale separation implies that the asymptotic limit $\epsilon = \lambda_{\rm D} / L \rightarrow 0$ is useful to solve for the plasma-sheath system.
In this limit, the Bohm criterion must be satisfied at the sheath entrance. 
A new derivation of the kinetic criterion, admitting a general ion velocity distribution, is presented.
It is proven that for $\epsilon \rightarrow 0$ the distribution of the velocity component normal to the target, $v_x$, and its first derivative must vanish for $|v_x| \rightarrow 0$ at the sheath entrance.
These two conditions can be subsumed into a third integral one after it is integrated by parts twice.
A subsequent interchange of the limits $\epsilon \rightarrow 0$ and $|v_x| \rightarrow 0$ is invalid, leading to a divergence which underlies the misconception that the criterion gives undue importance to slow ions.
\end{abstract}

\section{Introduction}

For a quasineutral plasma to exist next to a wall or a solid target, the most mobile charged species, typically electrons due to their smaller mass, must be reflected by the wall to achieve no net charge loss (and thus preserve quasineutrality).
The wall is thus negatively charged and the region where electron reflection occurs is called the `sheath' \citep{Langmuir-1923, Langmuir-1929, Tonks-1929}.
The sheath is positively charged and shields the quasineutral plasma from the negative charge on the wall. 
Its characteristic length scale is the Debye length, defined as $\lambda_{\rm D} = \sqrt{\epsilon_0 T_{\rm e} / n_{\rm e} e^2}$, where $n_{\rm e}$ is the electron density, $e$ is the proton charge, $T_{\rm e}$ is the electron temperature (in units of energy) and $\epsilon_0$ is the permittivity of free space.
The characteristic length scale $L$ of unmagnetised plasmas, or magnetised plasmas where the magnetic field is perpendicular to the target, is defined to be the smallest one between: the collisional mean free path of ions, the ionization mean free path of neutrals, and the target curvature \citep{Riemann-1991}.
It is usually much larger than the Debye length, such that the sheath is thin compared to the plasma
\begin{align} \label{ordering}
\epsilon = \frac{\lambda_{\rm D}}{L} \ll 1 \rm .
\end{align}
In magnetised plasmas with an oblique magnetic field impinging on the target, $L$ can also be the ion sound Larmor radius $\rho_{\rm S} = \sqrt{m_{\rm i}(ZT_{\rm e} + T_{\rm i})}/(ZeB)$, where $T_{\rm i}$ is the ion temperature, $B$ is the magnetic field strength, $m_{\rm i}$ is the ion mass and $Z$ is the ionic charge state \citep{Chodura-1982}\footnote{Strongly magnetised plasmas may also be solved in the limit $\rho_{\rm S} / L \rightarrow 0$. If the magnetic field is obliquely incident on the target, the `magnetised plasma sheath' comprises two regions: the charged `Debye sheath' of width $\sim \lambda_{\rm D}$ and a (usually larger) quasineutral `magnetic presheath' of width $\sim \rho_{\rm S}$ \citep{Geraldini-2024-Chodura}.}.
Once they reach the target, ions and electrons are absorbed, eventually recombine and are subsequently re-emitted as neutrals that penetrate back through the sheath all the way into the plasma before re-ionizing.


The physical principle treated in this paper states that the ions must reach the sheath at a minimum velocity, as first formalised by \cite{Bohm-1949} assuming mono-energetic (cold) ions:
\begin{align} \label{cold-Bohm}
- u_{x} \geqslant v_{\rm B} \equiv \sqrt{ZT_{\rm e} / m_{\rm i}} \rm .
\end{align}
Here $u_{x} = \vec{u} \cdot \vec{e}_x$, where $\vec{u}$ is the fluid velocity, $\hat{\vec{e}}_x$ is a unit vector normal to the target plane and pointing away from it, $v_{\rm B} = \sqrt{ZT_{\rm e}/m_{\rm i}}$ is the Bohm speed (or cold-ion sound speed) and $m_{\rm i}$ is the ion mass.
The negative sign in the velocity component is related to the choice of placing the target at $x=0$ and the plasma at $x > 0$.
The inequality (\ref{cold-Bohm}) came to be known as the `Bohm criterion'. 
It holds at the `sheath entrance' (or `sheath edge'), a position at this stage loosely defined as the distance from the wall below which collisions, ionisation, target curvature and curvature of ion Larmor orbits are negligible, and above which the space charge is negligible.
It ensures an increasingly positive space charge in the sheath as the wall is approached, consistent with an increasing electric field directed towards the wall to reflect electrons.

Since mono-energetic ions are a significant idealisation, a \textit{kinetic} formulation of the Bohm criterion, valid for arbitrary ion velocity distributions, is desirable.
This was first given by \cite{Harrison-Thompson-1959},
\begin{align} \label{kin-Bohm-orig}
v_{\rm B}^2 \int_{-\infty}^0 \frac{f(v_x)}{v_x^2} dv_x \leqslant \frac{n_{\rm e}}{Z} \rm .
\end{align}
Here $f$ is the ion distribution function of the velocity component normal to the target, $v_x = \vec{v}\cdot \hat{\vec{e}}_x$ with velocity $\vec{v}$, at the sheath entrance.
By plasma quasineutrality at the sheath entrance, the electron density $n_{\rm e}$ satisfies
\begin{align} \label{ne}
n_{\rm e}  = Z\int_{-\infty}^{0} f(v_x) dv_x \rm .
\end{align}
The distribution $f$ is, in practice, the full three-dimensional distribution function which has been integrated over the velocity components tangential to the target.
Note that the fluid criterion (\ref{cold-Bohm}) is \emph{always} fulfilled by the flow velocity $u_x$ defined by
\begin{align}
u_x \int_{-\infty}^0 f(v_x) dv_x = \int_{-\infty}^0 f(v_x) v_x dv_x \rm ,
\end{align}
provided that $f$ satisfies the kinetic criterion (\ref{kin-Bohm-orig}), as can be shown by two applications of Schwarz's inequality \citep{Harrison-Thompson-1959, Riemann-1991}.
In deriving (\ref{kin-Bohm-orig}), it has been assumed that no ions travel away from the target at the sheath entrance, such that $f(v_x) = 0$ for $v_x > 0$.

Although Harrison \& Thompson's kinetic criterion (\ref{kin-Bohm-orig}) stood the test of time, \cite{Hall-1962} promptly criticised their derivation for implicitly assuming the conditions $f(0^-) \equiv \lim_{v_x \rightarrow 0^{-}} f(v_x) = 0$ and $f'(0^-) = 0$ without justification.
The derivative of any function $G(\zeta)$ of a single variable $\zeta$ is denoted $G'(\zeta) \equiv dG(\zeta)/d\zeta$.
Performing a calculation that will be shown here to be flawed, Hall refutes $f'(0^-) = 0$ 
and concludes that (\ref{kin-Bohm-orig}) ascribes ``undue importance'' to the slow ions with $v_x \approx 0$. 

In parallel, \cite{Caruso-Cavaliere-1962} recognised the asymptotic framework of the plasma-sheath transition as a singular perturbation theory (see section 7.2 of \cite{Bender-Orszag}), following work by \cite{Bertotti-1961}.
In the limit $\epsilon \rightarrow 0$ the equations on the plasma and sheath scales are distinct, and thus the two regions can be treated separately.
The sheath entrance is defined in this limit by $x/L  \rightarrow 0$ and $x/\lambda_{\rm D}  \rightarrow \infty$ \citep{Bertotti-1965}.
In practice, any point $x_{\epsilon} = \epsilon^q L$ on an intermediate length scale with $q \in (0,1)$ satisfies $\lim_{\epsilon \rightarrow 0} x_{\epsilon} / \lambda_{\rm D} \rightarrow \infty$ and $\lim_{\epsilon \rightarrow 0}  x_{\epsilon}/L \rightarrow 0$ and is thus a valid reference choice for the sheath entrance\footnote{For small but finite $\epsilon$, the plasma and sheath solutions can be asymptotically matched to the solution obtained in a transition region on a suitably-chosen intermediate length scale between plasma and sheath \citep{Franklin-Ockendon-1970, Slemrod-2002, Riemann-2003, Riemann-2006, Riemann-1997, Riemann-2005}; one may therefore argue that this transition region is the most accurate manifestation of the concept of a sheath entrance.}.
The actual distribution function at this point is denoted $f_{\epsilon}(v_x)$, and the asymptotic sheath entrance distribution function is defined by
\begin{align} \label{f-epsilonzero}
f(v_x) \equiv \lim_{\epsilon \rightarrow 0} f_{\epsilon}(v_x) \rm .
\end{align}
The Bohm criterion at the sheath entrance arises as a necessary condition for an electron-repelling \emph{sheath} solution to exist in the limit $\epsilon \rightarrow 0$.

In section 3.2 of his seminal review, \cite{Riemann-1991} derived the kinetic Bohm criterion (\ref{kin-Bohm-orig}) in the limit $\epsilon \rightarrow 0$, including ion reflection and re-emission from the wall.
For comparison with this work, his derivation is considered for a perfectly absorbing wall.
There, it is proven in an intermediate step that $f(0^-) = 0$, while the condition $f'(0^-) = 0$ emerges from the assumption that $f$ can be represented by a Taylor expansion in kinetic energy $v_x^2 / 2$ near $v_x = 0$.
This assumption was criticised \citep{Fernsler-2005, Baalrud-2011, Baalrud-2012, Baalrud-2015} for restricting the class of possible ion distribution functions at the sheath entrance.
Another objection to (\ref{kin-Bohm-orig}) emphasises the divergence of its left hand side when applied to the actual distribution function,
\begin{align} \label{Bohm-problem}
\lim_{\epsilon \rightarrow 0} v_{\rm B}^2 \int_{-\infty}^0 \frac{ f_{\epsilon}(v_x) }{v_x^2} dv_x = \lim_{\epsilon \rightarrow 0} \infty \rm ,
\end{align} 
caused by $f_{\epsilon}(0) \neq 0$ and $f_{\epsilon}'(0) \neq 0$ for finite $\epsilon$.
Echoing \cite{Hall-1962} around 50 years later, \cite{Baalrud-2011, Baalrud-2012} claim that the kinetic criterion places ``undue importance'' to slow ions.
While this assertion appears reasonable in light of (\ref{Bohm-problem}), it is motivated by the misguided expectation that (\ref{kin-Bohm-orig}) should still apply \textit{as it is} to $f_{\epsilon}$ for small but finite $\epsilon$ \citep{Riemann-2012}.

In the remainder of this article, the derivation of the kinetic Bohm criterion (\ref{kin-Bohm-orig}) is generalised to remove the assumption, and prove the physical legitimacy of, $f'(0^-) = 0$. 
Condition (\ref{kin-Bohm-orig}) is shown to emerge from
\begin{align}
 \lim_{v_x \rightarrow 0^-} \lim_{\epsilon \rightarrow 0} f_{\epsilon} (v_x) \equiv f(0^-) = 0 \rm , \label{kin-Bohm-compa} 
 \end{align}
 \begin{align} \label{kin-Bohm-compb} 
 \lim_{v_x \rightarrow 0^-} \lim_{\epsilon \rightarrow 0} f_{\epsilon}' (v_x) \equiv f'(0^-) = 0 \rm , 
 \end{align}
 \begin{align}
\lim_{v_x \rightarrow 0^-} \lim_{\epsilon \rightarrow 0} \int_{-\infty}^{v_x}  f_{\epsilon}''(v_x') \ln (-1/v_x') dv_x' \equiv \int_{-\infty}^0  f''(v_x) \ln (-1/v_x) dv_x  \leqslant \frac{n_{\rm e}}{Z v_{\rm B}^2} \equiv \lim_{\epsilon \rightarrow 0} \frac{n_{\rm e, \epsilon}}{Z v_{\rm B}^2} \label{kin-Bohm} \rm ,
 \end{align}
by integrating (\ref{kin-Bohm}) by parts and imposing (\ref{kin-Bohm-compa})-(\ref{kin-Bohm-compb}).
In (\ref{kin-Bohm}), the electron density at $x_{\epsilon}$ was defined as $n_{\rm e, \epsilon}$. 
To conclude, (\ref{kin-Bohm-compa})-(\ref{kin-Bohm}) are combined to obtain the appropriate reformulation of (\ref{kin-Bohm-orig}) in terms of $f_{\epsilon}$.
It is explained that the divergence (\ref{Bohm-problem}) follows from an incautious interchange of the limits $\epsilon \rightarrow 0$ and $v_x \rightarrow 0^-$. 

\section{Derivation of the kinetic Bohm criterion}

In this section, the kinetic Bohm criterion is derived by imposing the requirement of an electron-repelling potential solution in the sheath near the sheath entrance.
In section~\ref{subsec-Poisson}, the collisionless ion kinetic equation in the sheath is solved and Poisson's equation is thus derived (assuming adiabatic electrons).
Then, in section 2.2 the ion density is expanded near the sheath entrance using a matched asymptotic expansion \citep{Bender-Orszag}.
Here, the only assumption on the form of the ion distribution function at the sheath entrance near $v_x= 0$ is that it may be expanded as a power series in $v_x$. 
In section 2.3, the charge density is evaluated for all possible values of the exponent related to the largest term in this power series.
Imposing an electron-repelling potential near the sheath entrance is shown in section~\ref{subsec-kinBohm} to lead to conditions (\ref{kin-Bohm-compa})-(\ref{kin-Bohm-compb}) and to the inequality form of (\ref{kin-Bohm}) and (\ref{kin-Bohm-orig}).
Finally, in section~\ref{subsec-higher} we verify that the equality form of the kinetic Bohm criterion (\ref{kin-Bohm-orig}) is consistent with an electron-repelling sheath solution.

\subsection{Poisson's equation in the sheath} \label{subsec-Poisson}

The following dimensionless variables are introduced: the sheath-scale position $\xi = x / \lambda_{\rm D}$, the (negative of the) electrostatic potential relative to the sheath entrance $\chi = \lim_{\epsilon \rightarrow 0} e(\phi(x_{\epsilon}) - \phi(x))/T_{\rm e}$, the ion velocity component directed towards the wall $w = -v_x / v_{\rm B}$, and the ion distribution function $g(w) = Zf (v_x) v_{\rm B} /n_{\rm e}$ satisfying $\int_{0}^{\infty} g \left( w \right) dw = 1$ from (\ref{ne}).

For $\epsilon \rightarrow 0$, in the sheath $x \leqslant x_{\epsilon}$, corresponding to $\xi \in [0, \infty]$, the ion distribution function $g_{\xi}(\xi, w)$ satisfies the kinetic equation 
\begin{align} \label{kineq}
w \frac{\partial g_\xi}{\partial \xi} + \chi' \frac{\partial g_\xi}{\partial w} = 0 \rm .
\end{align}
Equation (\ref{kineq}) results from the normalised full kinetic equation which has been integrated in the other two velocity components, after all the terms small in $\epsilon$ have vanished in the limit $\epsilon \rightarrow 0$.
The $\epsilon$-dependent terms are related to the target curvature, collisions and ionisation in the plasma, the magnetic force on the ions, and explicit time dependence on the plasma scale $v_{\rm B} / L = \epsilon v_{\rm B} / \lambda_{\rm D}$.
By imposing the boundary condition at the sheath entrance $g_{\xi}(\infty, w) = g \left( w \right)$, assuming a perfectly absorbing wall, $g_{\xi}(0, w<0) = 0$, and assuming no ion reflection within the sheath, $g_{\xi}(\xi, w<0) = 0$, one obtains $g_{\xi}(\xi, w) = g ( \sqrt{w^2 - 2\chi} )$ from equation (\ref{kineq}).
In order for no ions to be reflected, $\chi (\xi) \geqslant 0$ is required. 
For the electrons, a Boltzmann distribution is assumed, which is justified when the sheath reflects most electrons back into the bulk such that $e^{-\chi(0)} \ll 1$. 
Hence, Poisson's equation in the sheath is
\begin{align} \label{Poisson}
\chi''(\xi) = \int_{\sqrt{2\chi}}^{\infty} g \left( \sqrt{w^2 - 2\chi} \right) dw - e^{-\chi} \rm .
\end{align}

\subsection{Matched asymptotic expansion of the ion density near the sheath entrance} \label{subsec-matched}

In order to solve (\ref{Poisson}) for $\chi (\xi)$ locally near the sheath entrance (where $\chi = 0$), the electron density may be expanded as a Taylor series in $\chi \ll 1$, $e^{-\chi} = 1 - \chi + \frac{1}{2}\chi^2 + O(\chi^3)$. 
The ion density integral for $\chi \ll 1$ is calculated via a matched asymptotic expansion (see section 7.4 in \cite{Bender-Orszag}) hinging on the observation that the function $g ( \sqrt{w^2 - 2\chi} )$ can be represented using two different approximations with a common range of validity.
For $\bar w = \sqrt{w^2 - 2\chi} \ll 1$ (slow ions), the function $g(\bar w)$ is expanded as a power series close to $\bar w=0$,
\begin{align} \label{gapprox1}
g(\bar w) = \sum_{\nu=1}^{N_{3+}-1} g_{p_\nu} \bar w^{p_\nu} +  O(w^{p_{N_{3+}}}) \text{ for } \bar w \ll 1 \rm ,
\end{align}
where $p \equiv~ p_1 > -1$ (the density should remain finite) is defined to be the smallest power in the expansion of $g(\bar w)$ near $\bar w = 0$, $g_p \equiv g_{p_1} > 0$ is the corresponding positive constant coefficient ($g(w) > 0$), $p_{\nu} > p_{\nu-1}$ is the $\nu$th power in the expansion for $\nu > 1$, and $N_{3+}$ is the index corresponding to the smallest exponent in the expansion satisfying $p_{N_{3+}} > 3$.
By convention, it is considered that $g_{p'} = 0$ for $p'$ such that $p_{\nu} \neq p'$ for all possible values of the index $\nu$.
The expansion in (\ref{gapprox1}) is truncated such that terms $\ll \bar w^3$ are not retained, as these terms will be negligible throughout the subsequent analysis.
Note that the form of the distribution function $g(w)$ near $w=0$ assumed here is much more general than in previous derivations of the kinetic Bohm criterion.
For $\bar w \sim w \gg \sqrt{2\chi}$, $g ( \sqrt{w^2 - 2\chi} )$ can be Taylor expanded near $g(w)$,
\begin{align} \label{gapprox2}
g\left( \sqrt{w^2 - 2\chi} \right) = & g(w) - \chi \frac{g'(w)}{w} - \frac{1}{2} \chi^2 \frac{g'(w)}{w^3} + \frac{1}{2} \chi^2 \frac{g''(w)}{w^2} \nonumber \\
& + O\left( \frac{ \chi^3 g'(w) }{ w^5}, \frac{ \chi^3 g''(w) }{ w^4}, \frac{ \chi^3 g'''(w) }{ w^3} \right) \rm .
\end{align}

The two approximations (\ref{gapprox1})-(\ref{gapprox2}) have a common range of validity $\sqrt{2\chi} \ll w \ll 1$.
Therefore, (\ref{Poisson}) can be re-expressed by choosing a cutoff velocity parameter $w_{\rm c}$, generally satisfying $\sqrt{2\chi} \ll w_{\rm c} \ll 1$, and using the approximation (\ref{gapprox1}) for $w \leqslant w_{\rm c}$ and (\ref{gapprox2}) for $w \geqslant w_{\rm c}$.
This results in
\begin{align} \label{Poisson-perturbed}
\chi'' = ~ & Q_0 + Q_1 + Q_2 + \sum_{\nu=1}^{N_{3+}-1} \bar Q_{(p_\nu+1)/2} + O( \chi^{3} w_{\rm c}^{p-5}, \chi^3 , \chi w_{\rm c}^{p_{N_{3+}}-1}) = O(\chi^{(p+1)/2}, \chi) \rm ,
\end{align}
with the zeroth order, first order, second order, and slow ion charge densities, respectively, defined by:
\begin{align}
Q_0 = \int_{0}^{\infty} g \left( w \right) dw - 1 = 0 \rm , \label{Poisson-0}
\end{align}
\begin{align} 
Q_1 = \chi \left[ 1 - \int_{w_{\rm c}}^{\infty}  \frac{g' \left( w \right)}{w}  dw \right] = O(\chi, \chi w_{\rm c}^{p-1}) \rm , \label{Q1} 
\end{align}
\begin{align} 
Q_2 =  \frac{1}{2}  \chi^2 \left[   \int_{w_{\rm c}}^{\infty}  \frac{g'' \left( w \right)}{w^2}  dw -  \int_{w_{\rm c}}^{\infty}  \frac{g' \left( w \right)}{w^3}  dw - 1 \right] = O(\chi^2, \chi^2 w_{\rm c}^{p-3})  \rm ,  \label{Q2} 
\end{align}
\begin{align} \label{Qslow}
\bar Q_{(p_{\nu}+1)/2} & = g_{p_{\nu}} \left[ \int_{\sqrt{2\chi}}^{w_{\rm c}} \left(w^2 - 2\chi \right)^{p_{\nu}/2}  dw -  \int_0^{w_{\rm c}} w^{p_{\nu}} dw  \right] \nonumber \\
& =   g_{p_{\nu}} \left[ - \frac{ (2\chi)^{(p_{\nu}+1)/2} }{p_{\nu}+1} + \sum_{n=1}^{\infty} \frac{(2\chi)^n }{n!} \prod_{m=0}^{n-1} \left( m- \frac{p_{\nu}}{2} \right) \int_{\sqrt{2\chi}}^{w_{\rm c}} \frac{dw}{w^{2n-p_{\nu}}}  \right] \nonumber \\
& = O(\chi^{(p_\nu+1)/2}, \chi w_{\rm c}^{p_\nu - 1} ) \rm .
\end{align}
The quasineutrality condition at the sheath entrance (where $\chi = 0$ and $\chi''=0$) was used in (\ref{Poisson-0}).
The errors $O(\chi^{3} w_{\rm c}^{p-5}, \chi^3)$ in (\ref{Poisson-perturbed}) result from integrating the errors in (\ref{gapprox2}) in the interval $w\in [w_{\rm c}, \infty]$ and from another $O(\chi^3)$ term in the expansion of the electron density.
The error $O( \chi w_{\rm c}^{p_{N_{3+}}-1})$ comes from the term $\bar Q_{(p_{N_{3+}}+1)/2}$ which has been neglected in (\ref{Poisson-perturbed}).
The terms $Q_2$ and $\bar Q_{(p_{\nu}+1)/2}$ for $1<p_\nu \leqslant 3$ ($N_{1+} \leqslant \nu < N_{3+}-1$) are retained in equation (\ref{Poisson-perturbed}), despite being $\ll \chi$, in anticipation of a higher-order analysis that will be necessary for $p>1$ (see section~\ref{subsec-higher}).
\cite{Hall-1962} uses a Taylor expansion instead of a power expansion in (\ref{gapprox1}), thus considering $p=0$ (and $p_\nu =\nu-1$) or $p=1$ (and $p_\nu=\nu$) while keeping terms up to order $\sim \chi$, but incorrectly takes $w_{\rm c} = \sqrt{2\chi}$.
Even the more general expansion (\ref{Poisson-perturbed}) is invalid for $w_{\rm c} = \sqrt{2\chi}$ and $p\leqslant 1$, since some of the neglected terms are of order $O(\chi^{3}w_{\rm c}^{p-5})$, which would be $O(\chi^{(p+1)/2})$ just like the dominant terms in (\ref{Poisson-perturbed}).
For (\ref{Poisson-perturbed}) to have an error $\ll \chi$ when $p\leqslant 1$, it is required that $w_{\rm c}$ lie within the asymptotic region of overlap $\chi^{1/2} \lesssim \chi^{2/(5-p)} \ll w_{\rm c} \ll 1$.
In sections \ref{subsec-kinBohm} and \ref{subsec-higher}, it will be seen that the asymptotic region of overlap changes with the value of $p$ and with the order in $\chi$ of the expansion.
However, provided that such a region exists and $w_{\rm c}$ lies within it, the expansion up to the desired order is independent of the value of $w_{\rm c}$.

\subsection{Evaluation of the charge density} \label{subsec-charged}

The assumption of an electron-repelling sheath, $\chi(\xi) \geqslant \chi(\infty) = 0$, requires that the charge density $\chi''$ in equation (\ref{Poisson-perturbed}) become positive in the sheath.
The sign of $\chi''$ is established in the following cases sequentially: $-1< p < 1$, $p=1$, $1<p<3$, $p=3$, and $p>3$, thus covering all $p>-1$ in (\ref{gapprox1}).
It will be convenient to first evaluate the density contributions (\ref{Q1})-(\ref{Qslow}). 
Equations (\ref{Q1}) and (\ref{Q2}) for $Q_1$ and $Q_2$ can be evaluated to
\begin{gather} 
Q_1 =  \chi \left[ 1 - \int_0^{\infty} \left( g''(w) - \sum_{\nu = 1}^{N_{1-}} p_\nu (p_\nu-1) g_{p_\nu} w^{p_\nu-2} \right) \ln (1/w) dw \right]  \nonumber \\ 
 - \chi g_{1} \ln (1/w_{\rm c}) + \chi \sum_{\nu=1,~ p_\nu \neq 1}^{N_{3+}-1} \frac{p_\nu g_{p_{\nu}}}{p_{\nu}-1}  w_{\rm c}^{p_{\nu}-1} 
 + O(\chi w_{\rm c}^{p_{N_{3+}}-1})  \rm , \label{Q1-low} 
\end{gather}
\begin{gather} 
Q_2 = \frac{1}{4}  \chi^2 \left[ \int_{0}^{\infty}  \left( g'''' \left( w \right) - \sum_{\nu = 1}^{N_{3-}} g_{p_{\nu}} p_{\nu}(p_{\nu}-1)(p_{\nu}-2)(p_{\nu}-3) w^{p_\nu-4} \right) \ln (1/w)  dw  - 2  \right]  \nonumber \\
+ \frac{3}{4} \chi^2 g_{3} \left( 1 + 2\ln(1/w_{\rm c}) \right) -  \chi^2 \sum_{\nu = 1}^{N_{3-}} \frac{p_\nu (p_\nu -2)}{2(p_\nu -3)} g_{p_\nu} w_{\rm c}^{p_\nu -3} + O\left( \chi^2 w_{\rm c}^{p_{N_{3+}}-3} \right) \rm . \label{Q2-low}
\end{gather}
In (\ref{Q1-low}), $N_{1-}$ is the index corresponding to the largest exponent in the expansion satisfying $p_{N_{1-}} < 1$, while $N_{1+}$ is the index corresponding to the smallest exponent in the expansion satisfying $p_{N_{1+}} > 1$.
In (\ref{Q2-low}), $N_{3-}$ is the index corresponding to the largest exponent in the expansion satisfying $p_{N_{3-}} < 3$.
The derivation of (\ref{Q1-low}) and (\ref{Q2-low}) uses the following steps: (i) substitute $g(w) = g(w) - \sum_{\nu=1}^{N_{1-}} g_{p_\nu} w^{p_\nu} + \sum_{\nu=1}^{N_{1-}} g_{p_\nu} w^{p_\nu}$ in (\ref{Q1}) and $g(w) = g(w) - \sum_{\nu=1}^{N_{3-}} g_{p_\nu} w^{p_\nu} + \sum_{\nu=1}^{N_{3-}} g_{p_\nu} w^{p_\nu}$ in (\ref{Q2}); (ii) integrate explicitly the integrals involving $\sum g_{p_\nu} w^{p_\nu}$ to obtain the terms proportional to $1/ w_{\rm c}^a$ with $a>0$; (iii) integrate by parts (more than once for (\ref{Q2})) the integrals involving $g(w) - \sum g_{p_\nu} w^p$ and use (\ref{gapprox1}) to extract the boundary terms containing $\ln w_{\rm c}$ (present if $p_\nu = 1$ in (\ref{Q1-low}) or $p_\nu = 3$ in (\ref{Q2-low}) for some $\nu$); (iv) split the remaining integrals using $\int_{w_{\rm c}}^{\infty}dw (\ldots) = \int_{0}^{\infty}dw (\ldots) - \int_0^{w_{\rm c}} dw (\ldots)$ and substitute (\ref{gapprox1}) to evaluate the integrals $\int_0^{w_{\rm c}} dw (\ldots)$.
Equation (\ref{Qslow}) for $\bar Q_{(p_{\nu}+1)/2}$ becomes:
\begin{align} 
\bar Q_{(p_{\nu}+1)/2} = A_{(p_{\nu}+1)/2} \chi^{(p_{\nu}+1)/2} - \frac{p_{\nu} g_{p_{\nu}} \chi w_{\rm c}^{p_{\nu}-1}}{p_{\nu}-1} + \frac{p_{\nu}(p_{\nu}-2)}{2(p_{\nu}-3)} g_{p_{\nu}} \chi^2 w_{\rm c}^{p_{\nu}-3} + O(\chi^3 w_{\rm c}^{p_{\nu}-5}) \nonumber \\ \text{ for } p_{\nu} > -1, p_{\nu} \neq 2n-1 \text{ with } n \in \mathbb{N}^{\star} \rm , \label{Qslow-exp} 
\end{align}
\begin{align}
\bar Q_1 = - g_1\chi \left[ 1 + \frac{1}{2} \sum_{n=2}^{\infty} \frac{\prod_{m=1}^{n-1} \left( m - \frac{1}{2} \right)}{n!(n-1)} + \ln \left( \frac{w_{\rm c}}{\sqrt{2\chi}} \right) \right] 
+ \frac{1}{4} g_1 \chi^2 w_{\rm c}^{-2} + O(\chi^3 w_{\rm c}^{-4}) \rm , \label{Qslow-1} 
 \end{align}
 \begin{align}
 \bar Q_2 = g_3 \chi^2 \left[ 2 + 3 \sum_{n=3}^{\infty} \frac{\prod_{m=2}^{n-1} \left( m-\frac{3}{2} \right)}{n!(2n-4)}  + \frac{3}{2} \ln \left( \frac{w_{\rm c}}{\sqrt{2\chi}} \right) \right] - \frac{3}{2} g_3 \chi w_{\rm c}^2 + O(\chi^3 w_{\rm c}^{-2}) \rm , \label{Qslow-2}
 \end{align}
 where $A_{(p+1)/2}$ in (\ref{Qslow-exp}) is defined by
 \begin{gather}
 \frac{A_{(p_{\nu}+1)/2}}{2^{(p_{\nu}+1)/2}} = g_{p_{\nu}} \left[ \sum_{n=1}^{\infty} \frac{\prod_{m=0}^{n-1} \left( m - \frac{p_{\nu}}{2} \right)}{n! (2n-p_{\nu}-1)} - \frac{ 1 }{p_{\nu}+1} \right] \text{ for } p_{\nu}> -1, p_{\nu} \neq 2n-1 \text{ with } n \in \mathbb{N}^{\star} \rm .  \label{A-1p1}
  \end{gather}
Gauss' test can be used to show that the series in (\ref{Qslow-1})-(\ref{A-1p1}) are convergent for the given values of $p_{\nu}$.

\subsection{Electron-repelling potential in Poisson's equation: kinetic Bohm criterion} \label{subsec-kinBohm}

\begin{figure}
\centering
\includegraphics[scale=0.5]{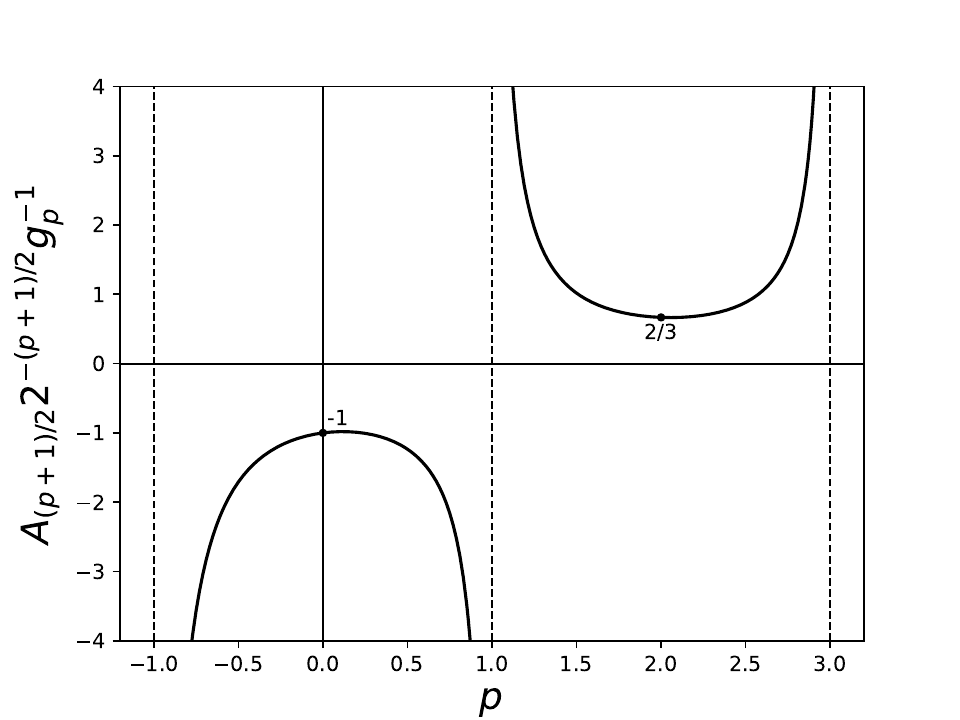}
\caption{The quantity $A_{(p+1)/2}/(2^{(p+1)/2} g_p)$ is plotted as a function of $p$ in the intervals $p \in (-1, 1)$ and $p \in (1, 3)$. The asymptotes at $p=-1$, $1$ and $3$ are shown as dashed lines. The value of the curve at $p=0$ and $2$ is marked.}
\label{fig}
\end{figure}

If $-1 < p< 1$, (\ref{Q1-low}) for $Q_1$, (\ref{Q2-low}) for $Q_2$, (\ref{Qslow-exp}) for $\bar Q_{(p+1)/2}$, and (\ref{Qslow-exp}), (\ref{Qslow-1}) or (\ref{Qslow-2}) for $\bar Q_{(p_\nu+1)/2}$ with $\nu \geqslant 2$ are inserted into (\ref{Poisson-perturbed}) to obtain
\begin{gather} \label{chi''-1p1}
\chi''(\xi) = A_{(p+1)/2} \chi^{(p+1)/2} + O(\chi^3 w_{\rm c}^{p-5}, ~\chi w_{\rm c}^{p_{N_{3+}}-1}, \chi^{(p_2+1)/2}) \rm .
\end{gather}
This equation is accurate for $ \chi^{1/2} \ll w_{\rm c} \lesssim 1 $, where the restriction $w_{\rm c} \ll 1$ is not necessary because the integral in (\ref{Poisson}) is dominated by the region near $w=\sqrt{2\chi}$ regardless of whether $g$ is approximated by (\ref{gapprox1}) or (\ref{gapprox2}) for $w \sim 1$.
Figure~\ref{fig} is a plot of the numerical evaluation of $A_{(p+1)/2}$ as a function of $p \in (-1,1) \cup (1, 3)$, illustrating that $ A_{(p+1)/2} < 0$ for $p \in (-1,1)$.
This assumed interval for $p$ has therefore led to a negative charge density, $\chi'' \leqslant 0$, in contradiction with an electron-repelling sheath.
Hence, the form (\ref{gapprox1}) requires $p \geqslant 1$: the distribution function satisfies $g(0^+) = 0$, corresponding to (\ref{kin-Bohm-compa}), and has a non-divergent (zero or finite) first derivative $g'(0^+)$.

If $p=1$, (\ref{Q1-low}) for $Q_1$, (\ref{Q2-low}) for $Q_2$, (\ref{Qslow-1}) for $\bar Q_{1}$, and (\ref{Qslow-exp}) or (\ref{Qslow-2}) for $\bar Q_{(p_\nu+1)/2}$, $\nu \geqslant 2$, are inserted into (\ref{Poisson-perturbed}) to obtain
\begin{align} 
\chi'' = & B_1 \chi \ln (1/\chi) + A_1 \chi + O(\chi^3 w_{\rm c}^{-4} , \chi w_{\rm c}^{p_{N_{3+}}-1}, \chi^{(p_2+1)/2})  \rm ,  \label{chi''p1}
\end{align}
where $A_1$ and $B_1$ are defined by
\begin{gather}
A_1 = g_1 \left[ \frac{1}{2}  \ln 2 - \frac{1}{2} \sum_{n=2}^{\infty} \frac{\prod_{m=1}^{n-1} \left( m - \frac{1}{2} \right)}{n!(n-1)} - 1 \right]  + 1 - \int_{0}^{\infty} g'' \left( w \right) \ln (1/w)  dw  \rm , \\
 B_1 = - g_1/2 < 0 \rm . \label{B1}
\end{gather}
Equation (\ref{chi''p1}) is accurate for $\chi^{1/2} \ll w_{\rm c} \ll 1$. This equation, which follows from the assumption $p=1$, is in contradiction with an electron-repelling sheath, since $B_1 < 0$ from (\ref{B1}) and $\ln (1/\chi) \gg 1$ for $\chi \ll 1$ imply that $\chi'' \leqslant 0$.
Hence, $p>1$ is required in (\ref{gapprox1}), corresponding to $g'(0^+) = 0$.
This conclusively proves that condition (\ref{kin-Bohm-compb}) must be satisfied at the sheath entrance. 

Considering $p>1$ and inserting (\ref{Q1-low}) for $Q_1$, (\ref{Q2-low}) for $Q_2$, and (\ref{Qslow-exp}) or (\ref{Qslow-2}) for $\bar Q_{(p_\nu+1)/2}$ in (\ref{Poisson-perturbed}) gives
\begin{align} \label{Poisson-chi}
\chi''(x) = & \chi \left[ 1 - \int_{0}^{\infty} g'' \left( w \right) \ln (1/w) dw \right]  + O( \chi^{(p+1)/2}, \chi^3 w_{\rm c}^{p-5}, \chi w_{\rm c}^{p_{N_{3+}}-1}, \chi^2 ) \rm .
\end{align}
Only when $p > 1$ does the asymptotic expansion of Poisson's equation in (\ref{Poisson-perturbed}) up to and including terms of order $\sim \chi$ become accurate for $\sqrt{2\chi} \leqslant w_{\rm c} \ll 1$, as seen from (\ref{Poisson-chi}) upon enforcing the constraint $w_{\rm c} \geqslant \sqrt{2\chi}$ arising from (\ref{Poisson}).
This is because the part of the integral in (\ref{Poisson}) close to $w=\sqrt{2\chi}$ is negligible compared to terms of order $\sim \chi$ irrespective of whether the approximation (\ref{gapprox1}) or (\ref{gapprox2}) is used there.
Hence, the expansion in \cite{Hall-1962}, which assumes $w_{\rm c} = \sqrt{2\chi}$, is correct only when the distribution function and its first derivative vanish at $w=0$, just like the expansion in \cite{Harrison-Thompson-1959} which was criticised by Hall.
For (\ref{Poisson-chi}) to be compatible with an electron-repelling sheath, the condition $\int_{0}^{\infty} \ln (1/w) g'' \left( w \right) dw < 1$ is required, which is the dimensionless form of (\ref{kin-Bohm}) without the equality.
In order to prove (\ref{kin-Bohm}), and consequently (\ref{kin-Bohm-orig}), it must still be verified that if the equality holds, 
\begin{align} \label{kin-g}
\int_{0}^{\infty} g'' \left( w \right) \ln (1/w) dw =  \int_{0}^{\infty} \frac{g \left( w \right)}{w^2} dw =  1 \rm ,
\end{align}
the higher-order terms in equation (\ref{Poisson-chi}) are consistent with an electron-repelling sheath solution with $\chi(\xi) \geqslant 0$.

\subsection{Higher-order analysis for marginally satisfied kinetic Bohm criterion} \label{subsec-higher}

If $1<p<3$, inserting (\ref{Q1-low}) for $Q_1$, (\ref{Q2-low}) for $Q_2$, (\ref{Qslow-exp}) for $\bar Q_{(p+1)/2}$, (\ref{Qslow-exp}) or (\ref{Qslow-2}) for $\bar Q_{(p_\nu+1)/2}$ with $\nu \geqslant 2$, and (\ref{kin-g}) in (\ref{Poisson-perturbed}) leads to
\begin{gather}
\chi'' = A_{(p+1)/2} \chi^{(p+1)/2} + O(\chi^3 w_{\rm c}^{p-5}, \chi w_{\rm c}^{p_{N_{3+}}-1}, \chi^{(p_2+1)/2}) \rm . \label{chi''1p3}
\end{gather}
This equation is accurate for $\chi^{1/2} \ll w_{\rm c} \ll \chi^{(p-1)/[2(p_{N_{3+}}-1)]}$.
From figure \ref{fig} it is seen that $A_{(p+1)/2} > 0$ for $1<p<3$, which makes (\ref{chi''1p3}) consistent with a positive space charge, $\chi'' \geqslant 0$.
If $p = 3$, inserting (\ref{Q1-low}) for $Q_1$, (\ref{Q2-low}) for $Q_2$, (\ref{Qslow-2}) for $\bar Q_{2}$, (\ref{Qslow-exp}) for $\bar Q_{(p_\nu+1)/2}$ with $\nu \geqslant 2$, and (\ref{kin-g}) into (\ref{Poisson-perturbed}) gives (note that $p=p_1 = 3$ implies $N_{3+} = 2$)
\begin{align} 
\chi'' = & B_2 \chi^2 \ln (1/\chi) + A_2 \chi^2 + O\left( \chi^3 w_{\rm c}^{-2}, \chi w_{\rm c}^{p_2-1} \right) \rm , \label{chi''p3} 
\end{align}
with
\begin{align}
A_2 = g_3 \left[ \frac{11}{4} + 3 \sum_{n=3}^{\infty} \frac{\prod_{m=2}^{n-1} \left( m-\frac{3}{2} \right)}{n!(2n-4)}  - \frac{3}{4} \ln 2  \right] + \frac{1}{4} \int_{0}^{\infty}  g'''' \left( w \right) \ln (1/w)  dw - \frac{1}{2}  \rm ,
\end{align}
\begin{align}
B_2 = 3g_3 / 4 > 0 \rm . \label{B2}
\end{align}
Equation (\ref{chi''p3}) is accurate for $\chi^{1/2} \ll w_{\rm c} \ll \chi^{1/(p_2-1)}$, with $p_2 > 3$, and is again consistent with a positive space charge, since $B_2 > 0$ from (\ref{B2}).
If $p>3$, inserting (\ref{Q1-low}) for $Q_1$, (\ref{Q2-low}) for $Q_2$, (\ref{Qslow-exp}) for $\bar Q_{(p_\nu+1)/2}$, and (\ref{kin-g}) into (\ref{Poisson-perturbed}) gives
\begin{align} \label{chi-highest}
\chi'' = A_2 \chi^2 + O\left( \chi w_{\rm c}^{p-1},  \chi^3 \right) \rm , 
\end{align}
\begin{align}
A_2 = \frac{1}{4} \int_{0}^{\infty} g'''' \left( w \right) \ln (1/w) dw - \frac{1}{2} =   \frac{3}{2}  \int_{0}^{\infty} \frac{g \left( w \right) }{ w^4 } dw - \frac{1}{2} > 0 \rm .  \label{A2}
\end{align}
Equation (\ref{chi-highest}) requires $\sqrt{2\chi} \leqslant w_{\rm c} \ll \chi^{1/(p-1)}$ to be accurate.
In (\ref{A2}), Schwartz's inequality and the equalities (\ref{Poisson-0}) and (\ref{kin-g}) constrain $\int_{0}^{\infty}  dw  g \left( w \right) / w^4  \geqslant 1$ \citep{Riemann-1991}, such that $A_2 > 0$.
Equation (\ref{chi-highest}) is compatible with an electron-repelling sheath, and exhausts all possible remaining values of $p$.
This completes the proof that an electron-repelling sheath solution (with adiabatic electrons) exists if (\ref{kin-Bohm-compa})-(\ref{kin-Bohm}), which can be combined into (\ref{kin-Bohm-orig}), are satisfied.

\section{Discussion and conclusion}

The physical interpretation of the kinetic Bohm criterion (\ref{kin-Bohm-orig}) in the limit $\epsilon \rightarrow 0$ is as follows.
When entering the sheath, slower ions undergo a much larger relative velocity increment than faster ions for a given potential drop, and by continuity they have a larger contribution to the ion density drop.
Hence, for mono-energetic ions there is a threshold velocity (\ref{cold-Bohm}) above which the ion density drop is less than the electron density drop, consistent with the positive space charge expected in the sheath.
Similarly, with an arbitrary velocity distribution instead of a mono-energetic one, conditions (\ref{kin-Bohm-compa})-(\ref{kin-Bohm-compb}) limit the number of slow ions with $v_x \approx 0$ to ensure that the ion density drop scales linearly with the potential drop, like the electron density drop, and not more strongly.
Condition (\ref{kin-Bohm-orig}) then results from requiring that the ion density drop be smaller than the electron one. 
At this point, it should be unsurprising that slow ions with $|v_x| \ll v_{\rm B}$ have a larger weight in this condition, forcing the bulk ions to compensate by moving faster towards the wall and causing the over-satisfaction of (\ref{cold-Bohm}).

If it is desirable to formulate (\ref{kin-Bohm-orig}) in terms of the actual distribution function $f_{\epsilon}$, this can be done carefully as follows.
Presuming that $f_\epsilon$ is, just like its limit $f$ in (\ref{f-epsilonzero}), differentiable at $v_x = 0$, the two conditions (\ref{kin-Bohm-compa}) and (\ref{kin-Bohm-compb}) imply that $f_{\epsilon}(0) \sim \epsilon^{r_1(q)} n_{\rm e} / Z v_{\rm B}$ and $f_{\epsilon}'(0) \sim \epsilon^{r_2(q)} n_{\rm e} / Z v_{\rm B}^2$, where $r_1(q) > 0$ and $r_2(q) > 0$ depend on $q$ in $x_{\epsilon} = \epsilon^q L$.
The small additional number of slow ions with $v_x \approx 0$ at $x_{\epsilon}$ is due to the replenishment of ions with $v_x \geqslant 0$ occurring in the small region $x \in [0, x_{\epsilon}]$, caused for example by rare (vanishing for $\epsilon \rightarrow 0$) collision or ionization events in this region. 
Integrating the third condition (\ref{kin-Bohm}) by parts twice gives 
\begin{align}
\lim_{v_x \rightarrow 0^-} \lim_{\epsilon \rightarrow 0} \left[  \int^{v_x}_{-\infty}  \frac{f_{\epsilon}(v_x')}{v_x'^2} dv_x' + f_{\epsilon}' (v_x) \ln (-1/v_x)   + \frac{f_{\epsilon}(v_x)}{v_x} \right]
 \leqslant \lim_{\epsilon \rightarrow 0} \frac{n_{\rm e, \epsilon}}{Z v_{\rm B}^2 } \label{lim-kin-Bohm-heart} \rm .
\end{align} 
If the limits on the left hand side of (\ref{lim-kin-Bohm-heart}) are interchanged, all terms diverge; 
without the interchange, the second and third terms vanish by (\ref{kin-Bohm-compa}) and (\ref{kin-Bohm-compb}), recovering (\ref{kin-Bohm-orig}).
The left hand side of (\ref{Bohm-problem}) results from interchanging the limits in (\ref{lim-kin-Bohm-heart}) without retrieving the additional terms that no longer vanish (and diverge) after the interchange.
The limits in (\ref{lim-kin-Bohm-heart}) can be combined without interchange by taking $v_x = - v_{\epsilon} $ with $v_{\epsilon} = \epsilon^r v_{\rm B}$ and $r \in (0,r_1(q))$, such that $\lim_{\epsilon \rightarrow 0} f_\epsilon(-v_\epsilon) / v_\epsilon = 0$ and $\lim_{\epsilon \rightarrow 0} f'_\epsilon(-v_\epsilon) \ln (1/v_\epsilon ) = 0$. 
This allows subsuming conditions (\ref{kin-Bohm-compa}), (\ref{kin-Bohm-compb}) and (\ref{lim-kin-Bohm-heart}) into the single condition\footnote{If $f_{\epsilon}$ is not differentiable at $v_x = 0$, such that $f_{\epsilon} (v_x) \sim \epsilon^{r_3} v_x^{p_{\epsilon}}$ for $v_x \ll v_{\rm B}$ with $r_3 > 0$ and $p_{\epsilon} \in (-1,0) \cup (0,1)$, the limits $\epsilon \rightarrow 0$ and $v_x \rightarrow 0_-$ are non-interchangeable in (\ref{kin-Bohm-compa}) (if $p_{\epsilon} \in (-1,0)$), (\ref{kin-Bohm-compb}) and (\ref{kin-Bohm}), and $r < r_3/(1-p_{\epsilon})$ is required for (\ref{lim-kin-Bohm-orig}) to hold.}
\begin{align}
 \lim_{\epsilon \rightarrow 0} \int^{-v_{\epsilon}}_{-\infty}  \frac{f_{\epsilon}(v_x)}{v_x^2} dv_x  \leqslant \lim_{\epsilon \rightarrow 0} \frac{n_{\rm e, \epsilon}}{Z v_{\rm B}^2 } \label{lim-kin-Bohm-orig} \rm  .
\end{align}
A small number (vanishing for $\epsilon \rightarrow 0$) of slow ions with $|v_x| < v_{\epsilon}$ are ignored in (\ref{lim-kin-Bohm-orig}) because their contribution to the density drop would be sharply overestimated due to the neglect of their replenishment in the region $x \leqslant x_{\epsilon}$.
Yet, the form of (\ref{lim-kin-Bohm-orig}) confirms the importance of slow ions with $|v_x| \ll v_{\rm B}$ in the kinetic Bohm criterion. 

~

This work has been carried out within the framework of the EUROfusion Consortium, via the Euratom Research and Training Programme (Grant Agreement No 101052200 — EUROfusion) and funded by the Swiss State Secretariat for Education, Research and Innovation (SERI). Views and opinions expressed are however those of the author(s) only and do not necessarily reflect those of the European Union, the European Commission, or SERI. Neither the European Union nor the European Commission nor SERI can be held responsible for them.

\appendix

\bibliography{Bohm}
\bibliographystyle{jpp}

\end{document}